\documentclass{ijcaArticle}
\usepackage{subcaption}
\usepackage{amsmath}
\usepackage[none]{hyphenat} 
\usepackage[sorting=none]{biblatex}

\bibliography{BibTex_Sample}

\ijcaVolume{VV}
\ijcaNumber{N}
\ijcaYear{YYYY}
\ijcaMonth{Month}

\ijcaVolume{*}
\ijcaNumber{*}
\ijcaYear{2018}
\ijcaMonth{--------}
\begin{document}

\title{Effectiveness of Artificial Intelligence in Stock Market Prediction Based on Machine Learning} 

\author{ 
   \large Sohrab Mokhtari \\[-3pt]
   \normalsize Electrical and Computer Engineering \\[-3pt]
    \normalsize Florida International University \\[-3pt]
    \normalsize Miami, USA. \\[-3pt]
    \normalsize	somokhta@fiu.edu \\[-3pt]
  \and
   \large Kang K Yen \\[-3pt]
   \normalsize Electrical and Computer Engineering \\[-3pt]
    \normalsize Florida International University \\[-3pt]
    \normalsize Miami, USA. \\[-3pt]
    \normalsize	yenk@fiu.edu \\[-3pt]
\and
   \large Jin Liu \\[-3pt]
   \normalsize Electrical and Computer Engineering \\[-3pt]
    \normalsize Florida International University \\[-3pt]
    \normalsize Miami, USA. \\[-3pt]
    \normalsize	jiliu@fiu.edu \\[-3pt]
}

\keywords{Machine learning, time series prediction, technical analysis, sentiment embedding, financial market.}

\maketitle
\begin{abstract} 
This paper tries to address the problem of stock market prediction leveraging artificial intelligence (AI) strategies. The stock market prediction can be modeled based on two principal analyses called technical and fundamental. In the technical analysis approach, the regression machine learning (ML) algorithms are employed to predict the stock price trend at the end of a business day based on the historical price data. In contrast, in the fundamental analysis, the classification ML algorithms are applied to classify the public sentiment based on news and social media. In the technical analysis, the historical price data is exploited from Yahoo Finance, and in fundamental analysis, public tweets on Twitter associated with the stock market are investigated to assess the impact of sentiments on the stock market's forecast. The results show a median performance, implying that with the current technology of AI, it is too soon to claim AI can beat the stock markets.
\end{abstract}
\section{Introduction}
Stock markets are always an attractive investment way to grow capital. With the development of communication technology, the stock markets are getting more popular among individual investors in recent decades. While year by year, the number of shareholders and companies is growing in the stock markets, many try to find a solution to predict a stock market's future trend. This is a challenging problem with a multitude of complex factors that are impacting the price changes. Here, prediction algorithms such as Kalman filter \cite{BFAUKF} and optimization methods such as Nash equilibrium \cite{mokhtari2021impact} can be helpful; but, for this specific problem, AI can play a significant role. For this, ML methods are developed in many research papers to evaluate the prediction power of AI in the stock markets. The ML algorithms that are implemented for this purpose mostly try to figure out patterns of data, measure the investment risk, or predict the investment future. \par 
This field's efforts have led to two central theoretical hypotheses: Efficient Market Hypothesis (EHM) and Adaptive Market Hypothesis (AMH). The EMH \cite{fama1991efficient} claims that the spot market price is ultimately a reaction to recently published news aggregation. Since news prediction is an impractical phenomenon, market prices are always following an unpredictable trend. This hypothesis implies that there is no possible solution to 'beat the market.' On the other hand, the AMH \cite{lo2004adaptive} is trying to find a correlation between the evidential EMH and the extravagant behavioral finance principles. Behavioral finance tries to describe the market trend by psychology-based theories. Regarding the AMH, investors can leverage the market efficiency weakness to gain profit from share trading.\par 
Relying on the AMH statement, there should be possible solutions to predict the future of market behavior. Considering this fact, along with the Dow theory \cite{kirkpatrick2019dow}, leads to the creation of two basic stock market analysis principles: fundamental and technical. Fundamental analysis tries to investigate a stock's intrinsic value by evaluating related factors such as the balance sheet, micro-economic indicators, and consumer behavior. Whenever the stock value computed by this strategy is higher/lower than the market price, investors are attracted to buy/sell it. On the other hand, the technical analysis only examines the stock's price history and makes the trading decisions based on the mathematical indicators exploited from the stock price. These indicators include relative strength index (RSI), moving average convergence/divergence (MACD), and money flow index (MFI) \cite{edwards2007technical}. \par 
Decades ago, the proposed market analysis was performed by financial analysts; but through the development of computing power and artificial intelligence, this process could also be done by data scientists. Nowadays, the power of ML strategies in addressing the stock market prediction problem is strengthening rapidly in both fundamental and technical analyses. In an early study leveraging ML for stock market prediction, Piotroski $et$ $al.$ \cite{piotroski2000value} introduced an ML model called F-Score to evaluate companies' actual share values. Their method was based on nine factors exploited by a company's financial reports divided into three main categories: profitability, liquidity, and operating efficiency. They implemented the F-Score algorithm on the historical companies' financial reports of the U.S. stock market for twenty years from 1976 to 1996 and presented remarkable outcomes. Some years later, Mohanram $et$ $al.$ \cite{mohanram2005separating}
proposed a developed ML algorithm named G-Score to decide on trading stocks. Their approach was based on fundamental analysis applying financial reports to evaluate three criteria: profitability, naive extrapolation, and accounting conservatism. They also showed their algorithm sufficiency by back-testing the U.S. stock market trend between 1978 and 2001. \par
Basically, in the fundamental analysis, unlike the technical analysis, the data is unstructured and hard to be processed for training an ML model. Nevertheless, many studies by leveraging this type of analysis proved it can lead to a rational prediction of the market price. Contrarily, to analyze the market based on a technical method, only historical price data is required. This data is a structured type of data that is directly available to the public. This resulted in a far higher volume of research papers studying the prediction of the stock market based on the technical analysis approaches. As one of the early studies in this field, at the beginning of the 90s, Kimoto $et$ $al.$ \cite{kimoto1990stock} worked on a feed-forward neural network (NN) algorithm \cite{surveyabbas} to predict the stock market exploiting the historical financial indicators such as interest rate, and foreign exchange rate. Their model was a decision-making tool in generating the signal of buying/selling shares. Although their model could be successful in the buy-and-hold strategy, it could not predict the signal of selling sufficiently. Therefore, other ML algorithms were examined to assess the prediction power of ML using technical data such as artificial neural network (ANN), random forest (RF), support vector machine (SVM), and naive Bayesian. In  \cite{patel2015predicting}, Patel $et$ $al.$ implemented four distinct ML algorithms on this problem, including artificial neural network (ANN), random forest (RF), support vector machine (SVM), and naive Bayesian. Ten-year-period testbed results clarified that the random forest algorithm can be more effective among other algorithms, especially while the input data is discretized. Moreover, in a very recent study, Zhong $et$ $al.$ \cite{zhong2019predicting} studied a comprehensive big data analytic procedure applying ML to predict a daily return in the stock market. They employed both deep neural network (DNN) and ANN to fit and predict sixty financial input features of the model. They concluded that the ANN performs better than the DNN, and applying principle component analysis (PCA) in the pre-processing step can improve prediction accuracy.
 \par 
This paper attempts to investigate the effectiveness of AI, particularly ML, in addressing stock market prediction. In this research, both technical and fundamental stock market analyses are applied to measure ML algorithms' accuracy in predicting market trends. Moreover, a multitude of ML algorithms such as logistic regression, $k$-nearest neighbor, random forest, decision tree, and ANN are employed to find the most accurate algorithms to be selected as a solution to this problem. The data used to generate ML models are acquired in real-time, and the purpose of this research is to evaluate the accuracy of a selling/buying/holding signal of a specific share for investors.\par 
The remainder of the paper is organized as follows: Section II is devoted to the problem description and motivations; Section III contains methodology; Section IV measures the performance of ML algorithms on a stock market testbed; Section V points out the conclusion and future works.
\section{Problem Description and Motivation}
Since the 90s, early studies attempted to predict the stock markets leveraging AI strategies. Many research studies are published to evaluate the performance of AI approaches in the stock market prediction. Researchers' enthusiasm in studying the stock market prediction problem is due to the tremendous daily volume of traded money in the stock markets.
\par
Generally, analyzing the stock markets is based on two primary strategies: technical analysis and fundamental analysis. In the first one, stockholders try to evaluate the stock markets regarding the historical price data and investigating the generated indicators exploited from this data, such as the RSI and the MACD. An ML model can do the same. It can be trained to find a logical pattern between the financial indicators and the stock's closing price. This can lead to a prediction model that estimates the stock price at the end of a business day. On the other hand, in the fundamental analysis, stockholders attempt to calculate an actual stock value based on its owner company's financial reports, such as the market cap or the dividends. If the estimated price value is higher than the stock price, the stockholders receive a selling signal, while if the estimated price is lower than the stock price, they receive a holding/buying signal. It is evident that any changes in a company's financial report can immediately affect the public sentiment on the news and social media. An ML model can investigate news and social media through the Internet to predict a positive/negative impact of the stock prices' fundamental indicators. Then, provide the action signal for the stockholders based on the public sentiment.\par
But, the question is how much these approaches can be effective in the prediction of the market. In other words, $"Can$ $AI$ $beat$ $the$ $stock$ $market?"$. This study is trying to employ the ML algorithms and evaluate their performance in predicting the stock market to answer this question. The regression models are employed to predict the stock closing prices, and the classification models are used to predict the action signal for stockholders. In the following section, the methodology for addressing this evaluation is explained.
\section{Methodology}
\label{sec:Methodology}
In this study, the stock market's prediction, leveraging ML tools, includes four main steps: dataset building, data engineering, model training, and prediction. This section is devoted to explaining each of these steps in detail.
\subsection{Dataset}
The first step of building an ML model is having access to a dataset. This dataset includes some features that train the ML model. The training procedure can be done with or without a set of labeled data called target values. If the training is based on a set of labeled data, the training procedure is called supervised learning; while, unsupervised learning does not need any target values and tries to find the hidden patterns in the training dataset.\par
In the problem of predicting the stock market, most datasets are labeled. For instance, the dataset includes some financial indicators such as RSI and MACD as features and the stock's closing price as the target value in the technical analysis approach. It is evident that the data associated with the technical analysis is continuous numbers, which is shown in time-series format data. On the other hand, in the fundamental analysis strategy, the features are some statements such as financial reports or investors' sentiments, and the target value is the signal of decision-making in buying/selling the stock. In this type of analysis, the data includes typically alphabetic inputs such as reports and sentiments.\par
Hopefully, most of the essential data required for this problem is available online such as historical stock prices or public sentiments in the news. The data employed in this study was acquired from two sources: technical analysis data available on $Yahoo$ $Finance$\footnote{https://finance.yahoo.com/}, and sentiment analysis data available on $Twitter$\footnote{https://twitter.com/}. The Yahoo Finance data includes the open, close, mid, high, low price, and volume values without missing samples; while, the Twitter dataset contains tweets from the public comprising news agencies and individuals that can have missing samples.
\begin{figure*}
    \centering
    \includegraphics[width=13cm]{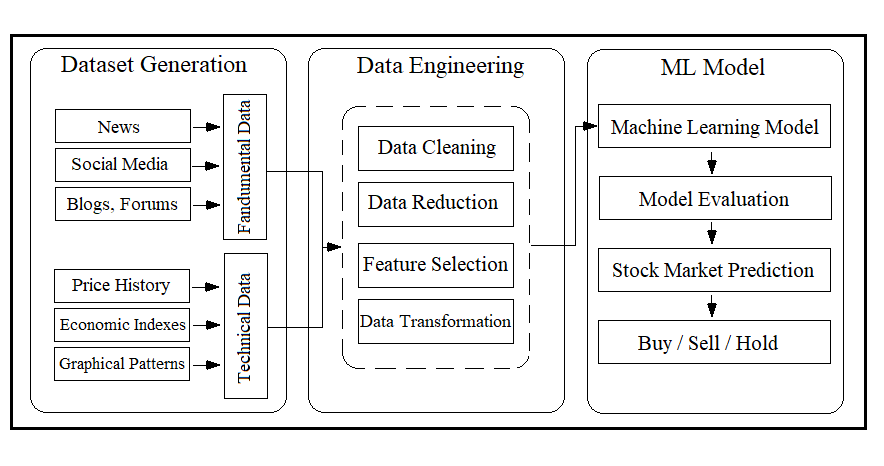}
    \caption{The framework of model training to predict the stock market.}
    \label{pic1}
\end{figure*}
\subsection{Data Engineering}
The data obtained from the proposed datasets requires to be pre-processed before being exploited in model training. Either technical analysis or fundamental analysis has several indicators applied in the model training step, and the most significant ones are explained in the following.
\subsubsection{Technical Analysis}
The historical stock prices are used to calculate appropriate financial indicators such as simple moving average (SMA), exponential moving average (EMA), RSI, MACD, and on-balance-volume (OBV) to build the input features of an ML training model. These indicators are explained in the following.\par
\textbf{SMA}. This indicator is the average of the most recent closing prices of a stock in a particular period. The mathematical calculation of the SMA is shown as below:
\begin{equation}
    SMA(t, N) = {\sum_{k=1}^{N}} \frac{CP(t-k)}{N}
\label{SMA}
\end{equation}
where $CP$ is the closing price, $N$ indicates the number of days that the $CP$ is evaluated, and $k$ shows the days associated with a particular $CP$.\par
\textbf{EMA}. This indicator tracks a stock price the same as the SMA, but it pays more attention to the recent closing prices by weighting them. Equation \ref{EMA} indicates the weighting process of this indicator. 
\begin{equation} \label{EMA}
    \begin{array}{cc}
        EMA(t, \Delta)=(CP(t)-EMA(t-1)) * \Gamma+EMA(t-1)
                \vspace{.2 cm} \\
        \Gamma =\frac{2}{\Delta+1}, \quad \Delta=\text {Time period EMA}
    \end{array}
\end{equation}
where $t$ is the present day, $\Delta$ is the number of days, and $\Gamma$ is the smoothing factor.\par
\textbf{MACD}. This indicator tries to compare the short-term and the long-term trends of a stock price. Equation \ref{MACD} describes this indicator as follow:
\begin{equation} \label{MACD}
    MACD = EMA(t, k) -  EMA(t, d)
\end{equation}
where $k$ and $d$ are the periods of short-term and long-term trends. Normally, these values are considered as $k=12$ and $d=16$ days.\par
\textbf{OBV}. This indicator uses a stock volume flow to show the price trend and indicates whether this volume is flowing in or out. The following equation explains the OBV concept:
\begin{equation}
OBV=OBV_{pr}+\left
    \{\begin{array}{ll}
    \text { volume,} & \text { if $CP$ }>\text { $CP$ }_{pr} \\
    0, & \text { if $CP$ }=\text { $CP$ }_{pr} \\
    -\text {volume}, & \text { if $CP$ }<\text { $CP$ }_{pr}
    \end{array}\right.
\end{equation}
where $OBV_{pr}$ is the previous OBV, $volume$ is the latest trading volume amount, and $CP_{pr}$ is the previous closing price.\par
\textbf{RSI}. This indicator is measuring the oversold or the overbought characteristic of a stock. Indeed, it shows the trend of buying/selling a stock. The RSI is described as:
\begin{equation} 
    \label{RSI}
    RSI = \frac{100}{1+RS(t)}, \quad RS(t)= \frac{AvgGain(t)}{AvgLoss(t)}
\end{equation}
where $RS(t)$ shows the rate of profitability of stock, $AvgGain(t)$ is the average gained profit of stock at time $t$, and $AvgLoss(t)$ indicates the average loss on that price. \par
\subsubsection{Fundamental Analysis}
Due to the unstructured nature of fundamental indicators, extracting data for fundamental analysis is not easy. However, the development of AI makes it possible to exploit data from the Internet for this purpose, leading to a more accurate stock market prediction. This data can be information related to the financial report of a company or the sentiment of investors. Literally, companies' financial reports instantly impact public sentiment and present themselves on social media, particularly Twitter. Thus, one way of evaluating the impact of fundamental data on market trends is by looking at public tweets. This strategy is called sentiment analysis of the stock market. In the sentiment analysis, the input data for training a model is basically unstructured, imported as text format to the model. The target of fundamental datasets is a binary value indicating the text's positive/negative impact on a specific stock. \par
Besides, based on the types of data, the pre-processing step differs. In the technical analysis, due to the data's numeric nature, it is essential to normalize the data before employing them for model training. The data normalization step is significant when the ML model wants to find a logical pattern in the input data. If the data are not on the same scale, the prediction process would not accurately perform. Thus, many functions are applied to normalize the data, such as MinMaxScaler, StandardScaler, and RobustScaler. In this paper, MinMaxScaler is used to scale the data and is described as below:
\begin{equation}
    a_{scaled}^{m}=\frac{(a_{i}^{m}-a_{min})}{(a_{max}-a_{min})}
    \label{minmax}
\end{equation}
where $a_{i}^{m}$ is the $i^{th}$ feature (indicator) from $m^{th}$ experiment (time sample), $a_{min}$ and $a_{max}$  are the minimum and the maximum values of the feature among the experiments, respectively. Also, $a_{scaled}^{m}$ indicates the scaled value for the $i^{th}$ feature of $m^{th}$ experiment.\par
On the other hand, in the fundamental analysis, the data is not numeric. The goal is to investigate the impact of a sentence --that can be a tweet on Twitter-- on public sentiment. Whenever using non-numerical data in training an ML model, the input data should be translated into numeric data. Thus, one way to do so is data labeling. 
\par
Feature selection means finding the most valuable features that lead to a more accurate ML model in a fewer computation time. This technique can be classified as a filter, wrapper, embedded, and hybrid methods \cite{chandrashekar2014survey}. In the filter method, correlation criterion  plays a significant role. Correlation is a measure of the linear relationship between two or more parameters. In this method, features showing the most correlation with the target are selected to build the model. Furthermore, to avoid redundant computation, the selected features should not be highly correlated to each other. To do so, the Pearson correlation technique is one of the most useful methods, which is described as below:
\begin{equation}
    Corr(i)=\frac{cov\left(a_{i}, b\right)}{\sqrt{var\left(a_{i}\right) * var(b)}}
    \label{correlation}
\end{equation}
where $a_{i}$ is the $i^{th}$ feature, $b$ is the target label, $cov()$ and $var()$ represent the covariance and the variance functions, respectively. The processed data could be employed to train the ML model, as shown in Fig. \ref{pic1}.
\subsection{Machine Learning Model Training}
Many ML algorithms have been employed to predict stock markets in research studies. Basically, there are two main categories of models to address this problem: classification models that try to help the investors in the decision-making process of buying, selling, or holding stock, and regression models that attempt to predict stock price movements such as the closing price of a stock. In research studies, over 90\% of the algorithms leveraged in predicting the stock market are classification models \cite{thanh2016building}. However, few studies tried to predict the exact stock prices using the regression models \cite{kazem2013support,yang2002support,efendi2018new}.\par
Among ML algorithms, the decision tree (DT), support vector machine (SVM), and artificial neural networks (ANN) are the most popular ones employed to predict stock markets \cite{nti2019systematic}. In this study, besides using the ANN, DT, and SVM models, logistic regression (LR), Gaussian naive Bayes (GNB), Bernoulli Naive Bayes (BNB), random forest (RF), $k$-nearest neighbor (KNN), and XGboost (XGB) are employed for classification strategy; moreover, linear regression and long short-term memory (LSTM) are used in regression problems. In the following, these algorithms are briefly explained.
\par
\textbf{ANN.} Originally came from the concepts in biology and consisted of various processing elements called neurons. These inter-connected neurons' task is majorly summing up the values of input parameters corresponding to their specified weights and then adding a bias. The number of neurons in the input should equal the number of neurons in the output. In the end, the output values are calculated after the transfer function is applied.\par
\textbf{DT.} Decision tree owns a structure similar to a tree, where each branch represents the test outcome, and each leaf indicates a class label. The structure also includes internal nodes, which represent the test on a particular attribute. The outcome is a final decision that provides the best fitting of calculated attributes of the best class. \par
\textbf{SVM.}  In the SVM model, examples are mapped as separated points in the space as vast as possible concerning each other. Hence, the predicted examples are also mapped to the same space and then categorized.\par
\textbf{LR.} Logistic Regression algorithm is one of the most suitable algorithms in regression analysis, especially when the dependent variable is binary, where a logistic function is leveraged for modeling.\par
\textbf{GNB, and BNB.} Gaussian Naive Bayes and Bernoulli Naive Bayes are considered supervised learning algorithms which are simple but very functional. Gaussian Naive Bayes includes prior and posterior probabilities of the dataset classes, while Bernoulli Naive Bayes only applies to data with binary-valued variables. \par
\textbf{RF.} Random forest algorithm includes a series of decision trees whose objective is to generate an uncorrelated group of trees whose prediction is more accurate than any single tree in the group.\par
\textbf{KNN.} The KNN is a well-known algorithm for classification problems, in which test data is used to determine what an unclassified point should be classified as. Manhattan distance and Euclidean distance are the methods that are used in this algorithm to measure the distance of the unclassified point to its similar points.\par
\textbf{XGB.}  A popular and open-source version of the gradient boosted trees algorithm, XGBoost is a supervised learning algorithm for the accurate prediction of an aimed variable based on its simpler and weaker models estimation.\par
\textbf{Linear Regression.} A subset of supervised learning, Linear Regression, is basically a first-order prediction, e.g., a line or a plane that best fits the dataset's data points. Any new point as the prediction will be located on that line or plane.\par
\textbf{LSTM.} Unlike standard feed-forward neural networks, the Long Term Short Memory algorithm owns feedback connections and is utilized in deep learning. This algorithm is widely used to classify problems and make predictions based on data in the time domain.\par
All the proposed algorithms are used to perform a stock market prediction, and their performance is compared to evaluate the sufficiency of ML in this problem.
The following subsection explains the metrics that are applied in the comparison procedure.
\subsection{Model Evaluation Metrics}
All prediction models require some evaluation metrics to investigates their accuracy in the prediction procedure. In ML algorithms, a multitude of metrics are available to measure the models' performance, including confusion matrix, and receiver operator characteristic (ROC) curve for classification models; and R-squared, explanation variation,  mean absolute percentage error (MAPE), root mean squared error (RMSE), and mean absolute error (MAE) for regression \cite{mokhtariICS}. The rest of this subsection is devoted to explaining the concept of these metrics.\par
\subsubsection{Confusion matrix} 
\begin{figure}[h] 
    \centering
    \includegraphics[width=9 cm]{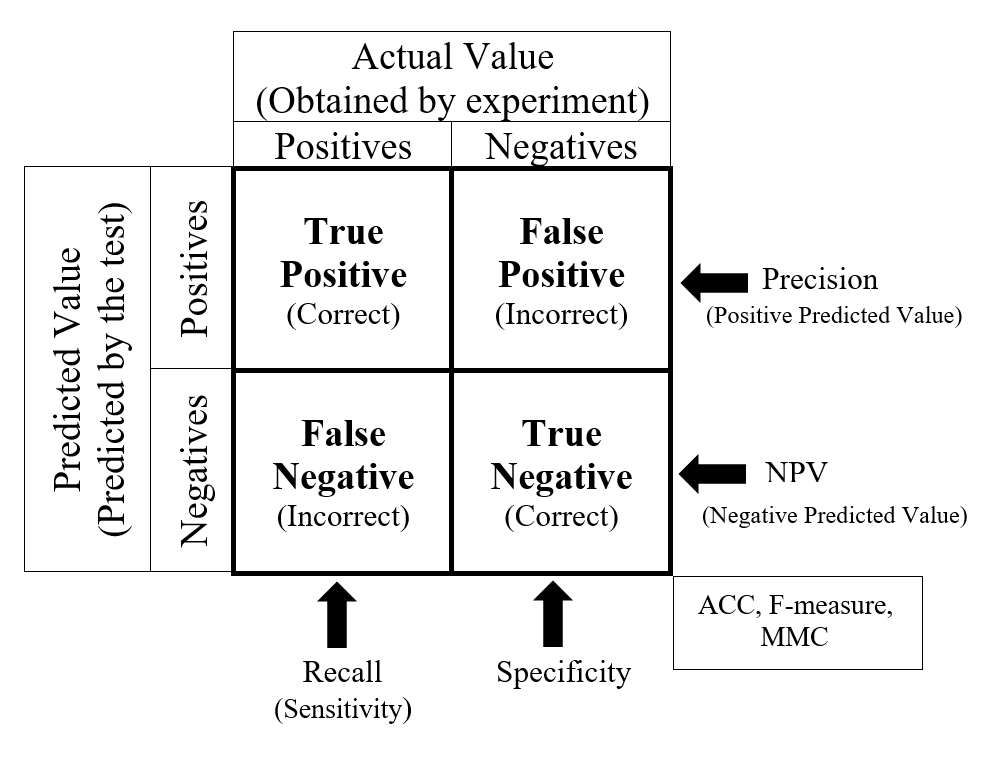}
    \caption{Confusion matrix explanation.}
   \label{confusionmatrix}
\end{figure}

This measure evaluates the accuracy of an ML model using a pre-known set of targeted data. Also, some other metrics, including sensitivity, specificity, precision, and F1-score, are resulted regarding this matrix. The sensitivity or recall is the likelihood of predicting true positive, while the specificity shows the true negative rate. Also, the precision indicates the accuracy of the true positive predicted classes. The F1-Score computes the balance between sensitivity and precision. Finally, the accuracy of the model would be the evaluation of the true predicted classes. Figure \ref{confusionmatrix} shows the confusion matrix concept. 
\subsubsection{ROC, AUC}
The receiver operator characteristic (ROC) curve includes two values: true-positive and false-positive rates. The ROC investigates the classifiers' performance among the whole range of class distributions and error costs. ROC curves are compared by the area under the curve (AUC) metric. The more values of AUC mention more accurate predicted outputs \cite{marzban2004roc}.\par
\subsubsection{R-squared ($R^{2}$)} The $R^{2}$ is a statistical measure indicating the variance portion for a dependent variable that's explained by an independent variable or variables in a regression model. It is also known as the coefficient of determination or the coefficient of multiple determination for multiple regression. Using regression analysis, higher $R^{2}$ is always better to explain changes in your outcome variable. If the R-squared value is less than 0.3, this value is generally considered a fragile effect size; if the R-squared value is between 0.3 and 0.5, this value is generally considered a low effect size; if the R-squared value is bigger than 0.7, this value is generally considered strong effect size. The following equation presents the formula for calculating the $R^2$ metric.
\begin{equation}
    R^{2}=1-\frac{\sum\left(y_{i}-\hat{y}_{i}\right)^{2}}{\sum\left(y_{i}-\bar{y}\right)^{2}}
\end{equation}
where $y_{i}$, and $\hat{y_{i}}$ are the $i^{th}$ actual and predicted value, respectively, and $\bar{y}$ shows the mean of actual values.
\subsubsection{Explanation Variation} The explained variance is used to measure the discrepancy between a model and actual data. In other words, it's the part of the model's total variance that is explained by factors that are actually present and are not due to error variance. The explained variation is the sum of the squared of the differences between each predicted value and the mean of actual values. Equation (\ref{EV}) shows the concept of explanation variation as below:
\begin{equation} \label{EV}
    EV =\sum(\hat{y_{i}}-\bar{y})^{2}
\end{equation}
where $EV$ is the explanation variation, $\hat{y_{i}}$ is the predicted value, and $\bar{y}$ indicates the mean of actual values.
\subsubsection{MAPE} The MAPE is how far the model’s predictions are off from their corresponding outputs on average. The MAPE is asymmetric and reports higher errors if the prediction is more than the actual value and lower errors when the prediction is less than the actual value. Equation (\ref{MAPE}) explains the mathematical formulation of this metric.
\begin{equation}\label{MAPE}
    MAPE=\frac{1}{n} \sum_{i=1}^{n}\left|\frac{y_{i}-\hat{y}_{i}}{y_{i}}\right|
\end{equation}
where $n$ is the number of experiments, $\hat{y_{i}}$ is the predicted value, and $y_{i}$ is the actual value for the $i^{th}$ experiment.
\subsubsection{RMSE} The computed standard deviation for prediction errors in an ML model is called RMSE. The prediction error or residual shows how far are the data from the regression line. Indeed, RMSE is a measure of how spread out these residuals are \cite{abedin2021bridge}. In other words, it shows how concentrated the data is around the line of best fit, as shown in Equation (\ref{RSME}). The smaller value of this metric represents a better prediction of the model.
\begin{equation} \label{RSME}
    RMSE=\sqrt{\frac{\sum_{i=1}^{n}\left(\hat{y}_{i}-y_{i}\right)^{2}}{n}}
\end{equation}
where $n$ is the number of experiments, $\hat{y_{i}}$ is the predicted value, and $y_{i}$ is the actual value for the $i^{th}$ experiment.\par
\subsubsection{MAE} The MAE is the sum of absolute differences between the target and the predicted variables. Thus, it evaluates the average magnitude of errors in a set of predictions without considering their directions. The smaller values of this metric mean a better prediction model. The following equation presents the mathematical MAE formula.
\begin{equation}\label{MAE}
    MAE=\frac{1}{n} \sum_{i=1}^{n}({y_{i}-\hat{y}_{i}})
\end{equation}
where $n$ is the number of experiments, $\hat{y_{i}}$ is the predicted value, and $y_{i}$ is the actual value for the $i^{th}$ experiment.\par
Regarding the proposed framework, the performance of ML algorithms on the prediction of stock markets can be evaluated. The following section implements ML algorithms on the real-life problem of the U.S. stock market prediction.
\begin{figure*}
    \begin{subfigure}{.5\textwidth}
      \centering
      \includegraphics[width=17cm]{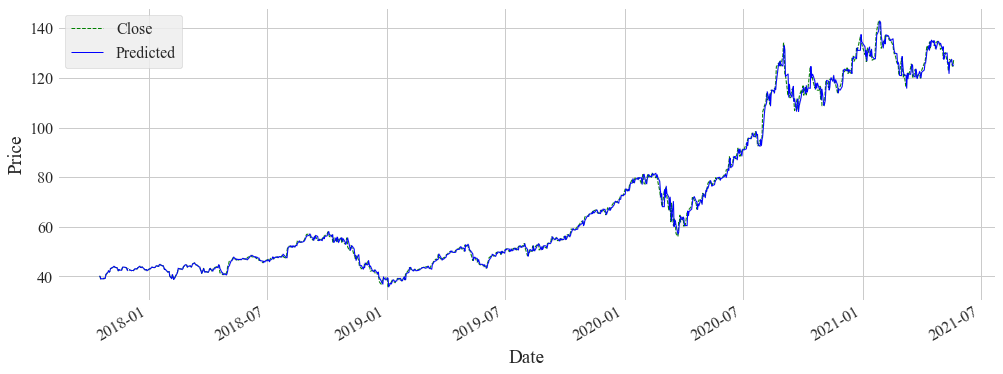}
      \caption{Prediction of the linear regression model.}
      \label{ROC1}
    \end{subfigure}%
\par
    \begin{subfigure}{.5\textwidth}
      \centering
      \includegraphics[width=17cm]{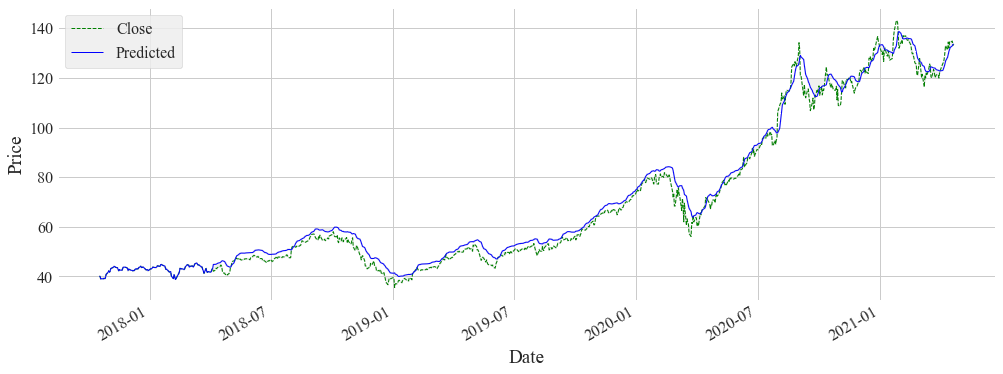}
      \caption{Prediction of the LSTM model.}
      \label{ROC2}
    \end{subfigure}
\caption{AAPL price prediction with the technical analysis approach.}
\label{price}
\end{figure*}
\section{Results and Discussion}
This section tries to illustrate the performance of the proposed methodology on the prediction of stock markets. For this, Python software is used to train the ML models and predict unforeseen data. First, the market prediction based on the technical analysis is evaluated, and then the fundamental analysis is investigated in this problem.\par
\begin{table}
    \begin{center}
    \caption{Models performance comparison, in technical analysis approach.}
        \begin{tabular}{|c||c|c|}
            \hline
            \textbf{Metric} & \textbf{Linear Regression} & \textbf{LSTM} \\
            \hline
            \hline
             \textbf{$R^{2}$} &  1.0 & 0.99  \\
             \hline
             \textbf{Explained Variation} & 1.0 & 0.99 \\
             \hline
             \textbf{MAPE} & 1.56 & 2.99 \\
             \hline
             \textbf{RMSE} & 1.82 & 3.42 \\
             \hline
             \textbf{MAE}  &  1.18 & 2.3  \\
             \hline
        \end{tabular}
        \label{metrics}
    \end{center}
\end{table}
\subsection{Technical Analysis Performance}
\begin{table*}
    \begin{center}
    \caption{Models performance comparison in fundamental analysis.}
        \begin{tabular}{|c||c|c|c|c|c|c|c|c|c|}
            \hline
             \textbf{Metrics}& \textbf{LR} & \textbf{GNB}  & \textbf{BNB} & \textbf{DT} & \textbf{RF} & \textbf{KNN} & \textbf{SVM} & \textbf{XGB}& \textbf{ANN}\\
            \hline
            \hline
             \textbf{Precision} & 0.729 & 0.636 &  0.644 &0.620&0.727&0.684&0.757&0.710&0.684\\
             \hline
             \textbf{Recall} & 0.727 & 0.634 &  0.644 &0.620&0.727&0.684&0.755&0.709&0.684\\
             \hline
             \textbf{F1-score} & 0.726 & 0.632  &  0.644 &0.620&0.727&0.684&0.755&0.709&0.684\\
             \hline
             \textbf{Accuracy} & 0.727 & 0.634 & 0.644 &0.620&0.727&0.684&\textbf{0.755}&0.709&0.684\\
             \hline
             \textbf{AUC} & 0.73 & 0.63 &  0.64 &0.62&0.73&0.68&0.76&0.71&0.68\\
             \hline
        \end{tabular}
        \label{fundCompare}
    \end{center}
\end{table*}
\begin{figure*}
    \begin{subfigure}{.5\textwidth}
      \centering
      \includegraphics[width=\linewidth]{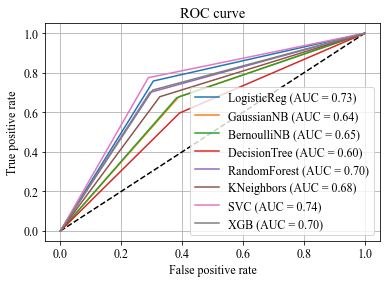}
      \caption{ROC curves.}
    \end{subfigure}
    \begin{subfigure}{.5\textwidth}
      \centering
      \includegraphics[width=\linewidth]{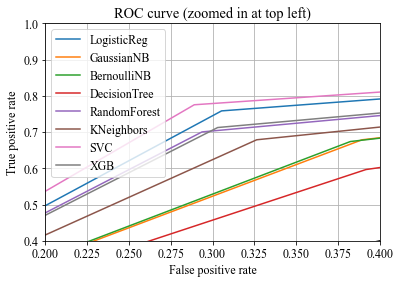}
      \caption{ROC curves from a closer view.}
    \end{subfigure}
\caption{ROC curves for classification algorithms.}
\label{ROC}
\end{figure*}
In this paper, the dataset for building a predictor model based on the technical analysis is exploited from the $"Yahoo$ $Finance"$ website. Indeed, it contains the historical data for a well-known stock called $AAPL$, which indicates Apple company information through a period of more than ten years between 2010 to 2021. The dataset includes 60 features such as open, high, low prices, the moving average, MACD, and RSI. The target is the close price, representing the final price of $AAPL$ at the end of a business day.\par
Then, the most correlated features to the target are selected, and then the redundant features that show a high correlation together are merged. Finally, the data is scaled by the MinMaxScaler function explained in Section 3.\par
The dataset is divided into three parts of the training data, validation data, and testing data to build the ML model. A large portion of the data is devoted to the training process, and the rest belongs to validation and testing. In the training process, the algorithm uses the training data to learn how to predict the target value accessible to the algorithm. Then, the model evaluates the performance of the prediction regarding the validation data. Finally, it can predict the unforeseen target of the testing dataset to compare with the true target values. In the end, by using the predicted and actual values of the closing price, the evaluation metrics can be measured. Table \ref{metrics} shows the comparison of the evaluation metrics. Moreover, Fig. \ref{price} shows the prediction of stock price based on the LR and the LSTM algorithms.\par
Regarding Table \ref{metrics}, the LR model is far better in predicting the $AAPL$ closing price compared to the LSTM model. Moreover, to illustrate the closing price's predicted and actual values, Fig. \ref{price} shows these values since 2018. The solid blue line shows the predicted value, and the dashed green line is the actual one.
\subsection{Fundamental Analysis Performance}
In this paper, a set of public tweets associated with $Apple$ company is employed to generate the required dataset available at \cite{yashchaudhary_2020}. In this case, the features are texts in $Twitter$, and the target is a binary value of impacted sentiment. If the content of the tweet has a positive impact on the stock market, the sentiment value would be 1, while a negative impact would give a -1 value to the sentiment. Then, the impacts of tweets on the specific stock are evaluated. Finally, the performance of ML in the prediction of buying/selling/holding signal is investigated.\par
The dataset includes nearly 6000 tweets, and the pre-processing data includes labeling the target values and employing the principle component analysis (PCA) \cite{howley2005effect} to reduce features dimension that show a high correlation.
Then, the proposed algorithms in Section 3 are used to classify the outcome of the model by negative or positive sentiment. Based on the evaluation metrics explained in the previous section, the performance of ML algorithms is compared and showed in Table \ref{fundCompare}. This table indicates that in this paper, the prediction of the public sentiment using ML algorithms does not show promising results. The most accurate algorithm is the SVM, with an accuracy of 76\%. Moreover, the performance of these algorithms is illustrated in Fig. \ref{ROC} that compares the ROC curves and also shows the AUC for each algorithm. In this figure, the SVM algorithm has the best AUC score.\par
\section{Conclusion}
This study tries to address the problem of stock market prediction leveraging ML algorithms. To do so, two main categories of stock market analysis (technical and fundamental) are considered. The performance of ML algorithms on the forecast of the stock market is investigated based on both of these categories. For this, labeled datasets are used to train the supervised learning algorithms, and evaluation metrics are employed to examine the accuracy of ML algorithms in the prediction process. The results show that the linear regression model predicts the closing price remarkably with a shallow error value in the technical analysis. Moreover, in the fundamental analysis, the SVM model can predict public sentiment with an accuracy of 76\%. These results imply that although AI can predict the stock price trends or public sentiment about the stock markets, its accuracy is not good enough. Furthermore, while the linear regression can predict the closing price with a sensible range of error, it cannot precisely predict the same value for the next business day. Thus, this model is not sufficient for long-term investments. On the other hand, the accuracy of classification algorithms in predicting buying, selling, or holding a stock is not satisfying enough and can result in loss of capital. \par
Nevertheless, many research studies on this topic are leveraging a hybrid model that employs both the technical analysis and the fundamental analysis in one ML model to compensate for the individual algorithms' downsides. This could increase the accuracy in the prediction process that implies an exciting topic for future studies. 
Based on this study, it seems that AI is not close to the prediction of the stock market with reliable accuracy. Maybe in the future, with AI development and especially computation power, a more precise model of stock market prediction can be available. Still, so far, there is no reputable model that can beat the stock market.
\printbibliography
\end{document}